\documentclass[a4paper,11pt]{article}

\usepackage{amsmath, amsthm, amsfonts, amssymb, bbm}
\usepackage[mathscr]{eucal}

\theoremstyle{plain}
\newtheorem{proposition}{Proposition}[section]

\theoremstyle{definition}

\theoremstyle{remark}
\newtheorem{remark}[proposition]{Remark}

\newcommand{\ud}{\mathrm{d}}

\newcommand{\skal}[2]{\langle #1 , #2 \rangle}

\newcommand{\id}{ \mathrm{id}  }

\newcommand{\V}[1]{\mathbf{#1}}

\newcommand{\order}{\mathcal{O}}

\newcommand{\schw}{\mathcal{S}}

\newcommand{\R}{\mathbb{R}}
\newcommand{\1}{\mathbbm{1}}

\newcommand{\F}{\mathcal{F}}

\newcommand{\tensor}{\otimes}
\newcommand{\betrag}[1]{\left| #1 \right|}

\begin{document}

\begin{flushright}
DESY 06-047 \\
ZMP-HH/06-05
\end{flushright}
\hfill
\begin{center}
{\LARGE Remarks on twisted noncommutative quantum field theory} \\
\hfill \\
{\large Jochen Zahn \\ II. Institut f\"ur Theoretische Physik, Universit\"at Hamburg \\ Luruper Chaussee 149, 22761 Hamburg, Germany \\ jochen.zahn@desy.de \\
\hfill \\
March 29, 2006 \\}
\hfill \\
\end{center}




\begin{abstract}
We review recent results on twisted noncommutative quantum field theory by embedding it into a general framework for the quantization of systems with a twisted symmetry. We discuss commutation relations in this setting and show that the twisted structure is so rigid that it is hard to derive any predictions, unless one gives up general principles of quantum theory. It is also shown that the twisted structure is not responsible for the presence or absence of UV/IR-mixing, as claimed in the literature.
\end{abstract}

\section{Introduction}

Noncommutative quantum field theory (NCQFT) is quantum field theory on the noncommutative Minkowski space, which is generated by coordinates $q^{\mu}$ subject to the commutation relations
\begin{equation*}
  [q^{\mu}, q^{\nu}] = i \sigma^{\mu \nu}.
\end{equation*}
Alternatively, it can be formulated via the $\star$-product of ordinary functions, see section~\ref{sec:setup}. The main motivations for the study of such models come from Gedankenexperiments on the possible localization of events~\cite{DFR} and the theory of open strings in the presence of a background $B$-field~\cite{SW}. For reviews see, e.g., \cite{Szabo, Bahns}.

Recently, it has been proposed to reconsider the question of violation of Lorentz invariance in NCQFTs. This was triggered by the realization that it is possible to twist the coproduct of the universal envelope $U \mathcal{P}$ of the Poincar\'e algebra such that it is compatible with the $\star$-product.  Already in \cite{Watts} it was shown that the $\star$-product naturally arises from a quasitriangular structure in the Hopf algebra corresponding to the translation group. Soon afterwards, it was shown that this quasitriangular structure is generated by a twist \cite{Oeckl}. 
Also the embedding into the (euclidean) Poincar\'e group was discussed there.
In~\cite{Chaichian, Wess} this was reformulated in dual language for the proper Poincar\'e algebra.
Subsequently, there have been claims about the violation of the Pauli principle \cite{Balachandran} and the absence of UV/IR-mixing \cite{BalachandranUVIR} in this twisted setting. Also an axiomatic characterization of twisted NCQFT was attempted \cite{Chaichian2}.

One aim of the present paper is to reach a more general understanding of QFT in the presence of a twisted symmetry. In this sense, it is related to the field of $q$-deformed quantum mechanics (see, e.g., \cite[Chapter 2]{ChaichianDemichev} for an overview) and the study of quantum systems with quantum symmetry (see, e.g., \cite{PW, Fiore, BraidedQFT}). We follow the philosophy outlined in \cite{WessEtAl}: Each time we encounter a bilinear map involving two spaces carrying a representation of the symmetry group we deform this map with the twist. The general setup is presented in section~\ref{sec:setup} and applied to NCQFT in section~\ref{sec:NCQFT}.
In section~\ref{sec:commutator} we consider two different commutators that appear naturally in the twisted setting. Unfortunately, both are lacking some crucial properties of the usual commutator.
Thus, we are leaving the safe grounds of established quantum mechanics.
This becomes even more evident in section~\ref{sec:time-evolution}. There it is shown that it is in general not possible to add a localized interaction (a source, for example) to the Hamiltonian without getting in serious trouble with the correspondence principle. We argue that this makes it extremely difficult to derive any predictions in the twisted setting (at least none that are not already present in conventional NCQFT). 
In section~\ref{sec:interactions} we make some remarks on the effect of the twist in the interacting case and in particular on the claim that the UV/IR-mixing is absent in the twisted setting~\cite{BalachandranUVIR}. We conclude with a summary.

\subsection*{Added Note}
While the present paper was written up, the preprint~\cite{Tureanu}, appeared. It has some overlap with the present paper, in particular it is in agreement with the results presented in section~\ref{sec:interactions}.

\section{Setup}
\label{sec:setup}

Let $\mu: \schw(\R^4) \otimes \schw(\R^4) \to \schw(\R^4)$ be the point-wise product of Schwartz functions. Then the $\star$-product can be defined as\footnote{Note that we use a different notation than in~\cite{WessEtAl}. Our $\F$ corresponds to $\F^{-1}$ there.} $\mu_{\star} = \mu \circ \F$ with
\begin{equation*}
  \F= e^{-\frac{i}{2} \sigma^{\mu \nu} P_{\mu} \otimes P_{\nu}}.
\end{equation*}
In~\cite{WessEtAl} the twist was interpreted as a formal power series in some deformation parameter. We will not do so in the present paper. Instead, we give it a rigorous meaning by going to momentum space.

The Poincar\'e algebra can be embedded into the Lie algebra $\Xi$ of vector fields and this in turn into the algebra $U \Xi$ that is obtained from the universal enveloping algebra by dividing out the ideal generated by the commutation relations in $\Xi$. Following \cite{WessEtAl}, one can equip $U \Xi$ with the structure of a Hopf algebra by defining the coproduct, counit and antipode through
\begin{align*}
  \Delta(u) =& u \otimes 1 + 1 \otimes u, & \Delta(1) =& 1 \otimes 1, \\
  \varepsilon(u)  = &0, & \varepsilon(1) = & 1, \\
  S(u)  = & -u, & S(1)  = & 1,
\end{align*} 
where $u \in \Xi$. This definition can be extended to $U \Xi$ by requiring $\Delta$ and $\varepsilon$ to be algebra homomorphisms and $S$ to be an antialgebra homomorphism. Furthermore, one can give $U \Xi$ a $*$-structure by defining
\begin{equation}
\label{eq:star-structure}
  u^*(f) = (S(u)(f^*))^*
\end{equation}
for $f \in \schw(\R^4)$ and $u \in \Xi$ and extending this as an antialgebra homomorphism. Furthermore, we note that $\F$ fulfills
\begin{align}
\label{eq:2cocycle}
  (\id \otimes \Delta)\F (\1 \otimes \F) = & (\Delta \otimes \id)\F (\F \otimes \1) \\
\label{eq:counital}
  (\varepsilon \otimes \id) \F = & \1 \\
\label{eq:real}
  (S \otimes S) (\F^{* \otimes *}) = & \F_{21}
\end{align}
where $\F_{21}$ is the transposed $\F$ (in our case $\F_{21} = \F^{-1}$). Thus, $\F^{-1}$ is a real (\ref{eq:real}), counital (\ref{eq:counital}) 2-coclycle (\ref{eq:2cocycle}), see, e.g., \cite{Majid}. From (\ref{eq:2cocycle}) it follows that the $\star$-product is associative and due to (\ref{eq:real}) it respects the $*$-structure\footnote{Note that the notations $\star$ and $\mu_\star$ are interchangeable in the present paper.}: $(f \star g)^* = g^* \star f^*$.

Now we consider the compatibility of the $\star$-product with Poincar\'e transformations.
Let $\xi \in U \mathcal{P}$. The point-wise product fulfills
\begin{equation}
\label{eq:a_mu}
 \xi \circ \mu = \mu \circ \Delta(\xi).
\end{equation} 
Here we identified $\xi$ with its action on $\schw(\R^4)$.
Now we want to find a deformed coproduct $\Delta_{\star}$ that fulfills $\xi \circ \mu_{\star} = \mu_{\star} \circ \Delta_{\star}(\xi)$. Using (\ref{eq:a_mu}), we find
\begin{equation*}
  \xi \circ \mu_{\star} = \mu \circ \Delta(\xi) \circ \F = \mu_{\star} \circ \F^{-1} \circ \Delta(\xi) \circ \F.
\end{equation*}
Thus, one defines
\begin{equation*}
 \Delta_{\star}(\xi) =  \F^{-1} \circ \Delta(\xi) \circ \F.
\end{equation*}
Since $\F \in U \mathcal{P} \otimes U \mathcal{P}$, it is clear that $\Delta_{\star}(\xi) \in U \mathcal{P} \otimes U \mathcal{P}$. In fact one explicitly finds \cite{Chaichian, Wess}
\begin{align}
\label{eq:DeltaP}
  \Delta_{\star}(P_{\mu}) = & P_{\mu} \otimes 1 + 1 \otimes P_{\mu} \\
  \Delta_{\star}(M_{\mu \nu}) = & M_{\mu \nu} \otimes 1 + 1 \otimes M_{\mu \nu} \nonumber \\
   - & \frac{1}{2} \sigma^{\alpha \beta} \left( g_{\mu \alpha} \left( P_{\nu} \otimes P_{\beta} -P_{\beta} \otimes P_{\nu} \right) - g_{\nu \alpha} \left( P_{\mu} \otimes P_{\beta} - P_{\beta} \otimes P_{\mu} \right) \right). \nonumber
\end{align}
There is a general theorem (see, e.g., \cite{Majid}, Thm 2.3.4) stating that $\Delta_\star$, together with $\epsilon_\star = \epsilon$ and $S_\star(\xi) = \chi^{-1} S(\xi) \chi$ with $\chi = S (\F_{(1)}) \F_{(2)}$ again define a Hopf algebra. Note that in our particular case $\chi = \1$, i.e., $S_\star = S$. From this and (\ref{eq:real}) it follows that we still have a Hopf $*$-algebra with the old $*$-structure (\cite{Majid}, Prop 2.3.7). Furthermore, one can show \cite{WessEtAl} that there is a triangular structure (or $R$-matrix) $R = \F_{21}^{-1} \F$. In our particular case we have
\begin{equation}
\label{eq:R}
  R = e^{i \sigma^{\mu \nu} P_{\mu} \otimes P_{\nu}}.
\end{equation}

Now suppose we are given vector spaces $A, B, C$ that carry a representation of the Poincar\'e algebra and a map $\nu: A \otimes B \to C$ that is compatible with this action. Then it is in the spirit of \cite{WessEtAl} to deform this map to $\nu_{\star} = \nu \circ \F$. As above, one then has $a \circ \nu_{\star} = \nu_{\star} \circ \Delta_{\star}(a)$. Now consider some special cases: 

\begin{itemize}
\item Let $A$ be an algebra carrying a representation of the Poincar\'e algebra and $\cdot$ the product $A \otimes A \to A$. Applying the above principle, one gets a new algebra $A_{\star}$, being identical to $A$ as a vector space, but with product $\star = \cdot \circ \F$. Due to (\ref{eq:2cocycle}) this product is associative. Note that if $A$ is a $*$-algebra, then due to (\ref{eq:real}), the new $\star$-product is compatible with the old $*$-structure: $(a \star b)^* = b^* \star a^*$.

\item Let $A$ be an algebra with a representation on a vector space $V$ and $\cdot : A \tensor V \to V$ be the corresponding left action. Applying the above principle, one first deforms $A$ to $A_{\star}$. Then one defines the action $\star : A_{\star} \otimes V \to V$ by $\star = \cdot \circ \F$. That this action defines a representation, i.e. $(a \star b) \star v = a \star (b \star v)$, follows again from (\ref{eq:2cocycle}).

\item If $V$ is even a Hilbert space, one should also define a new scalar product that is compatible with the adjoint in $A_{\star}$. The scalar product can be viewed as a bilinear map $\bar{V} \otimes V \to \mathbbm{C}$, where $\bar{V}$ is the conjugate vector space. The new scalar product $( \cdot , \cdot )_\star$ can then be defined in the obvious way. It remains to be shown that it is positive definite and compatible with the $*$-structure of $A_{\star}$. Note that in order to be consistent with~(\ref{eq:star-structure}), one defines the action of $U \Xi$ on $\bar{V}$ by $\xi \bar{v} = \overline{S(\xi^*) v}$. Also note that to the above left action of $A$ on $V$ there corresponds the right action $\bar{v} \cdot a = \overline{a^* \cdot v}$ on $\bar{V}$. This action can of course also be deformed in the obvious way. Due to~(\ref{eq:real}), we have $\bar{v} \star a = \overline{a^* \star v}$. The compatibility with the $*$-structure of $A_{\star}$, i.e. $(v, a \star w)_\star = (a^* \star v, w)_\star$, is now again a consequence of~(\ref{eq:2cocycle}). Unfortunately, there seems to be no general proof that $( \cdot, \cdot )_\star$ is positive definite\footnote{If one interprets $\F$ as a formal power series, one has positive definiteness in the sense of formal power series, since the  zeroth order component is the old one.}. Thus, this has to be checked explicitly in each example.

\end{itemize}

\section{The application to NCQFT}
\label{sec:NCQFT}

It is now fairly obvious how to apply the above to NCQFT. Identifying $A$ with the free field algebra and $V$ with the Fock space, we get a new product of quantum fields and a new action on the Fock space\footnote{Note that the action of the twist on tensor products of $L^2$-functions is well-defined in momentum space. Thus, we do not have to restrict to Schwartz functions, since no point-wise products are involved.}. It only remains to be checked that the new scalar product is positive definite. This is indeed the case, in fact it is the old one: Let $f \in L^2(\R^{3m}), g \in L^2(\R^{3n})$. Then
\begin{align*}
  (f, g)_{\star} = \delta_{m n} \frac{1}{m!} \sum_{\pi \in S_m}  \int \prod_{i=1}^m \frac{\ud^3 \V{k_i}}{2 \omega_i} \ & \bar{f}(\V{k_1}, \dots , \V{k_m}) g(\V{k_{\pi(1)}}, \dots , \V{k_{\pi(m)}} ) \\
   & \times e^{-\frac{i}{2} (-\sum_i k^+_i) \sigma (\sum_j k^+_j)}
\end{align*}
Here we used the notation $k^+ = (\omega_k, \V{k})$. Obviously, the twisting drops out. This is analogous to the property $\int \ud^4x \ f \star g(x) = \int \ud^4x \ f \cdot g(x)$ for test functions $f$ and $g$.

Obviously, the same construction can be done for a fermionic Fock space.

\begin{remark}
\label{rem:BDFP}
We remark that the new product of quantum field follows naturally from the smeared field operators introduced in \cite{BDFP}:
\begin{equation*}
  \phi_f(q) = \int \ud^4x \ \phi(q+x) f(x) = \int \ud^4k \ \hat{f}(k) \check{\phi}(k) \otimes e^{ikq}.
\end{equation*}
We then have
\begin{align*}
  \phi_f(q) \phi_g(q) = & \int \ud^4k_1 \ud^4k_2 \ e^{-\frac{i}{2} k_1 \sigma k_2}  \hat{f}(k_1) \hat{g}(k_2) \check{\phi}(k_1) \check{\phi}(k_2)  \otimes e^{i(k_1+k_2)q} \\
  = & \phi_{\F f \otimes g}^2(q).
\end{align*}
Here we used the notation
\begin{equation*}
  \phi^n_f(q) = \int \prod_{i=1}^n \ud^4k_i \ \hat{f}(k_1, \dots, k_n)  \prod_i \check{\phi}(k_i) \otimes e^{i(k_1 + \dots + k_n)q}.
\end{equation*}
Note that this notation deviates from the one used in \cite{BDFP}. In our notation one has the generalized formula
\begin{equation*}
  \phi^n_f(q) \phi^m_g(q) = \phi^{m+n}_{\F f \otimes g}(q).
\end{equation*}
Note that in \cite{BDFP} (and already in \cite{DFR}) quantum fields are elements of (or rather affiliated to) $\mathfrak{F} \otimes \mathcal{E}$, where $\mathfrak{F}$ is the field algebra and $\mathcal{E}$ the $C^*$-algebra generated by the quantum coordinates $q^{\mu}$.
An element of $\mathfrak{F}$ is then obtained by applying $\text{id} \otimes \omega$, where $\omega$ is a state on $\mathcal{E}$. Thus, the action of the field algebra on the Fock space is different than in the approach followed here.
\end{remark}

\begin{remark}
\label{rem:Balachandran}
In \cite{Balachandran} the twisted product was realized by a new definition of the creation and annihilation operators:
\begin{equation*}
  \tilde{a}(\V{k}) = a(\V{k}) e^{\frac{i}{2} k^+ \sigma P} , \quad \tilde{a}(\V{k})^* = a(\V{k})^* e^{-\frac{i}{2} k^+ \sigma P}.
\end{equation*}
Obviously, one then has $\tilde{a}(f) \Psi = a(f) \star \Psi$ for any Fock space vector $\Psi$ (and analogously for $a(f)^*$).
\end{remark}

\section{The twisted commutators}
\label{sec:commutator}

We turn to a question that is very important for finding a consistent interpretation of the new field algebra. Of course one is inclined to keep the interpretation of $\phi(f)$ as a field operator and of $a(f)^*, a(f)$ as creation and annihilation operators. But then they should fulfill some commutation relation that is compatible with the classical Poisson bracket. Since this classical bracket is not affected by the twist (at least if one uses Peierls definition, see~\cite{Diplom}), we would want the $\star$-commutator to give the usual result. This, however, is not the case for
\begin{align*}
  [\phi(f) \overset{\star}{,} \phi(g) ] = & \phi(f) \star \phi(g) - \phi(g) \star \phi(f) \\
  = & \int \ud^4k_1 \ud^4k_2 \ \hat{f}(k_1) \hat{g}(k_2) \\
  & \times \left( \check{\phi}(k_1) \check{\phi}(k_2) e^{- \frac{i}{2} k_1 \sigma k_2} - \check{\phi}(k_2) \check{\phi}(k_1) e^{\frac{i}{2} k_1 \sigma k_2} \right),
\end{align*}
as has already been noted in \cite[p.73f]{Bahns}. It is not even a $c$-number. But of course it fulfills the usual algebraic requirements antisymmetry, Leibniz rule and Jacobi identity.

\begin{remark}
This is the form of the commutator considered in \cite{Chaichian2} and denoted by $[\phi(f), \phi(g)]_{\star}$. Thus, our twisted NCQFT does not fulfill the locality axiom posed in there, even in the case of space-like noncommutativity.
\end{remark}

One can of course also consider the commutator as a map
\begin{align*}
  [ \cdot, \cdot ] : & \ A \otimes A \to A \\
  [ \cdot, \cdot ] : & \ a \otimes b \mapsto ab - ba.
\end{align*}
Then it is natural to deform it to the twisted commutator
\begin{equation*}
  [ \cdot, \cdot ]_{\star} = [ \cdot, \cdot ] \circ \F = \mu_{\star} - \mu_{\star} \circ R \circ \tau.
\end{equation*}
Here $\tau$ is the transposition and $R$ is the triangular structure (\ref{eq:R}). Note that this commutator was also used for a deformed Lie bracket of vector fields in \cite{WessEtAl}. In the context of NCQFT, it has already been proposed in \cite{Bahns} in the language of \cite{BDFP} (cf. remark \ref{rem:BDFP}). There, it simply amounts to postulating the commutator
\begin{equation*}
  [\phi \otimes f, \psi \otimes g] = [\phi, \psi] \otimes f g
\end{equation*}
for elements of $\mathfrak{F} \otimes \mathcal{E}$. It has already been remarked in \cite{Bahns} that it is neither antisymmetric nor fulfilling the Jacobi identity. However, it is possible to prove a Jacobi identity that involves the $R$-matrix \cite{WessEtAl}:
\begin{equation*}
  [a,[b,c]_\star]_\star = [[a,b]_\star , c]_\star + [R_{(1)} b, [ R_{(2)} a, c]_\star ]_\star.
\end{equation*}
There is an obvious similar formula expressing a twisted antisymmetry. While these formulae are general, there seems to be no analogous general formula for the Leibniz rule. In the concrete example of NCQFT, however, we have
\begin{equation*}
  [ a, b \star c]_\star = [a,b]_\star \star c + \F^{-2}_{(1)} b \star [ \F^{-2}_{(2)} a, c]_\star .
\end{equation*}
This can most elegantly be seen in the notation of~\cite{BDFP}.

We can now compute the twisted commutator of two fields:
\begin{align*}
  [\phi(f), \phi(g)]_{\star} = & \int \ud^4k_1 \ud^4k_2 \ e^{- \frac{i}{2} k_1 \sigma k_2} \hat{f}(k_1) \hat{g}(k_2) [\check{\phi}(k_1), \check{\phi}(k_2)] \\
  = & i \int \ud^4k_1 \ \hat{f}(k_1) \hat{g}(-k_1) \check{\Delta}(k_1).
\end{align*}
We see that the twisting drops out and we obtain the usual result. In particular, we have twisted commutativity if the supports of $f$ and $g$ are space-like separated. This seems to indicate that in the case of a twisted symmetry one should demand the correspondence principle between the classical Poisson bracket and the twisted commutator of the basic variables. We remark that Pusz and Woronowicz \cite{PW} found completely analogous twisted canonical commutation relations involving the $R$-matrix in a second quantization of a finite system with $SU_q(N)$-symmetry.
This may be seen as a hint that this is a general structure (see also \cite[Chapter 2]{ChaichianDemichev} and references therein). However, it should be noted that the vanishing of the commutator has a physical meaning, the possibility of simultaneous measurement. It is not clear wether the vanishing of the twisted commutator can be given a similar meaning.

It should also be noted that, as has already been remarked in \cite{Bahns}, the twisted commutator of products of fields does not coincide with the usual one, and does in particular not vanish for space-like separated supports. This is illustrated in the following example:
\begin{align*}
  [\phi^2(f_1 \otimes f_2), \phi(f_3)]_{\star} = & \int \left( \prod_{i=1}^3 \ud^4k_i \hat{f}_i(k_i) \right) e^{-\frac{i}{2} ( k_1 + k_2 ) \sigma k_3} [\check{\phi}(k_1) \check{\phi}(k_2), \check{\phi}(k_3) ] \\
  = & i \int \ud^4k \ \hat{f}_1(k) \check{\phi}(k) \int \ud^4p \ e^{\frac{i}{2} k \sigma p} \hat{f}_2(p) \hat{f}_3(-p) \check{\Delta}(p) \\
   + & i \int \ud^4k \ \hat{f}_2(k) \check{\phi}(k) \int \ud^4p \ e^{\frac{i}{2} k \sigma p} \hat{f}_1(p) \hat{f}_3(-p) \check{\Delta}(p).
\end{align*}
We emphasize once more that this is also in conflict with the correspondence principle.
Note that it does not help to use $\phi(f_1) \star \phi(f_2)$ instead.

\begin{remark}
\label{rem:formal}
If one interprets the twisting as a formal power series, then the $[ \cdot , \cdot ]_{\star}$-commutator is local at every order (this is not the case for $[ \cdot \overset{\star}{,} \cdot]$). If one identifies the scale of noncommutativity with the Planck scale, then $\sigma$ could be interpreted to be of $\order(\hbar)$. In this sense the twisting would yield higher powers of $\hbar$. It may be interesting to investigate this further. However, we will not do so in the present work.
\end{remark}

Thus, the upshot of this section is that we have two natural commutators in the twisted setting. The first one fulfills the usual algebraic requirements but deviates from the classical Poisson bracket. The other one does not have very nice algebraic properties, but at least reproduces the classical Poisson bracket for simple fields (but not for products of fields). In any case the correspondence principle has to be modified considerably, so one is leaving the safe grounds of established quantum mechanics.

\section{Time evolution}
\label{sec:time-evolution}

In ordinary quantum theory, the time evolution of observables is given by the commutator with the Hamiltonian $H$. If we want to keep this in the twisted setting, we have to decide which commutator to use. Due to~(\ref{eq:DeltaP}) one expects that the time-evolution fulfills the Leibniz rule, at least if $H$ is time-independent. Thus, one should use the $[ \cdot \overset{\star}{,} \cdot ]$-commutator. The classical equations of motion, however, do not change. Thus, the requirement that the time-evolution is, to zeroth order in $\hbar$, identical to the classical evolution leads to the condition $[H \overset{\star}{,}a] = [H,a] + \order (\hbar^2)$ for all observables $a$. But in the preceding section we have seen that the $[ \cdot \overset{\star}{,} \cdot ]$-commutator in general deviates from $[\cdot , \cdot]$ already at zeroth order\footnote{Even if one interprets the twist as a formal power series and assumes that $\sigma$ is of $\order(\hbar)$ (cf. remark~\ref{rem:formal}), then the two commutators still deviate at first order in $\hbar$.}. So the only general way to make things consistent seems to be to require that $H$ is invariant under the symmetry operation involved in the twist (in our case the translations), since then we have $[H \overset{\star}{,}a] = [H,a]$. But then the structure becomes very rigid, since a change of $H$ must be accompanied by a change of the twist. It is not even clear if there is such a new $\F$ in general.

\begin{remark}
There seems to be some similarity to the observation~\cite[Remark 2.2]{Bahns} that in the Hamiltonian approach to NCQFT the interacting Heisenberg field does not fulfill the equation of motion (see also~\cite{Heslop}). There, however, the problem appears only if the interaction is at least quadratic and if there is noncommutativity between space and time. Here, instead, the problem is connected only to translation invariance and thus already arises for a source term and also in the case of space-like noncommutativity.
\end{remark}

In the case when $H$ is not invariant under the symmetry involved in the twist, one could of course simply postulate the time-evolution $\dot{a} = i [H,a]$. But this time-evolution would in general be incompatible with the twisted algebra structure:
$[H,a \star b] \neq [H,a] \star b + a \star [H,b]$.
Thus, one would have to use the old algebra and nothing would have changed.

In order to see how this rigidity makes a meaningful exploration of the new framework impossible, consider the following example: Applying a creation operator $a(g)^*$ on the vacuum twice, one gets the two-particle wave function
\begin{equation*}
  \Psi_{\F g \otimes g}(\V{k_1}, \V{k_2}) = \sqrt{2} g(\V{k_1}) g(\V{k_2}) \cos \frac{k_1^+ \sigma k_2^+}{2}.  
\end{equation*}
Thus, the modulus $\betrag{ \Psi_{ \F g \otimes g }(\V{k_1}, \V{k_2})}$ is reduced for momenta $\V{k_1}, \V{k_2}$ such that $k_1^+ \sigma k_2^+ \sim 1$. This can of course only happen if $\Delta_i \Delta_j \sim \lambda_{nc}^{-2}$ for $i,j$ in noncommuting directions. Here $\Delta_i$ denotes the typical width of $g$ in the direction~$i$. In this sense the wave function $\Psi_{\F g \otimes g}$ is more narrow in momentum space and thus has a wider spread in position space in the noncommuting direction (of course the effect is tiny for realistic energies if $\lambda_{nc}$ is identified with the Planck length). Thus, one gets the impression that the twisting disfavors the occurrence of several particles with the same wave function if this wave function is simultaneously localized in noncommuting directions. If this was true, this might be an elegant resolution of the uncertainty problem posed in \cite{DFR}.

But the discussion above is of course at best heuristic. On the shaky ground we are exploring, we do not have any good intuition about what the repeated action of $a(g)^*$ might actually signify. And the two-particle wave function $\Psi_{g \otimes g}$ is still an element of our Fock space. Thus, we would like to make a statement in more operational terms. Now $\Psi_{g \otimes g}$ is, up to normalization, the two-particle component of the coherent state $e^{\lambda a(g)^*} \Omega$. This, in turn, can be characterized by being the ground state corresponding to the Hamiltonian
\begin{equation*}
  H = H_0 + \lambda ( a(f)^* + a(f) ),
\end{equation*}
where $H_0$ is the usual free Hamiltonian and $f = - P_0 g$. Taking this as a motivation, it would be interesting to find the ground state corresponding to this Hamiltonian in our twisted setting, i.e. the eigenvector $\Psi$ with the lowest eigenvalue $H \star \Psi = E \Psi$. This can be done, and it turns out that the corresponding two-particle wave function is indeed more narrow than $\Psi_{g \otimes g}$ (it is even more narrow than $\Psi_{\F g \otimes g}$). However, it is not clear if this result has any meaning, because $H$ is, in the twisted setting, not the generator of the time-evolution. In the present example, this is easily seen for the time-evolution of a field, as we already computed the $[ \cdot \overset{\star}{,} \cdot ]$-commutator of two fields in the preceding section.

\begin{remark}
In \cite{Balachandran} it has been claimed that in the twisted setting Paulis exclusion principle is no longer valid. The authors conclude this from the fact that in the case of twisted anticommutation relations one has in general $a(g)^* \star a(g)^* \neq 0$. But of course the fermionic wave functions are still antisymmetric. It is simply not clear what Paulis principle tells us in the twisted case (as in the example above, we do not know what the repeated action of $a(g)^*$ actually means). One should rather look for a statement in operational terms. First steps in this direction have been taken in~\cite{Chakraborty}.
However, in the light of the preceding discussion it is doubtful that this can be done consistently.
\end{remark}

\begin{remark}
\label{rem:YF}
In view of the problems discussed here, one might of course define the interacting directly by the equation of motion, i.e. use the Yang-Feldman formalism~\cite{YF}. In the context of NCQFT this has first been proposed in~\cite{Unitarity}. However, one has to bear in mind that if the interaction is not translation invariant, the interacting field will not transform covariantly under translations, i.e. $\phi_{int}(\tau_a f) \neq U_a \phi_{int}(f) U_a^{-1}$, where $\tau$ is the action on test functions and $U$ the Hilbert space representation of the translation group. This will make the $\star$-product of interacting fields more complicated.
\end{remark}

\section{Interactions}
\label{sec:interactions}

The effect of interactions can be studied by formally computing the $n$-point functions of the interacting field, defined, e.g., by the Yang-Feldman formalism. If the interaction is translation invariant, the interacting field transforms covariantly under translations (see remark~\ref{rem:YF}), and we have
\begin{multline*}
  \skal{\Omega}{\phi_{int}(f_1) \star \dots \star \phi_{int}(f_n) \Omega} \\ = \int \left( \prod_{i=1}^n \ud^4k_i \hat{f}_i(k_i) \right) e^{-\frac{i}{2} \sum_{i < j} k_i \sigma k_j } \skal{\Omega}{\check{\phi}_{int}(k_1) \dots \check{\phi}_{int}(k_n) \Omega}.
\end{multline*}
On the right hand side, all the loops are contained in the vacuum expectation value. Obviously, the twisting factor does not interfere at all with these and has no effect on the divergencies, and in particular does not influence the UV/IR-mixing. Thus the absence or presence of the UV/IR-mixing does only depend on the choice of the interaction term\footnote{This also seems to be at odds with the results of \cite{Oeckl}.}.

In \cite{BalachandranUVIR} the old (pointwise) product of fields was used\footnote{More precisely, the inverse $\star$-product between the operators $\tilde{a}, \tilde{a}^*$ (see remark \ref{rem:Balachandran}) was used. But since these operators already realize the $\star$-product, the combined effect amounts to the pointwise product.}. Thus, it is not surprising that the UV/IR-mixing is absent there.

If, however, the $\star$-product of fields is used for the interaction term, the UV/IR-mixing will be exactly as usual. In particular, one finds distorted dispersion relations \cite{BDFP, InPrep}.

\section{Summary}

We developed twisted NCQFT by embedding it in the general context of quantization of systems with twisted symmetries.
We discussed two different commutators. One of them fulfilled the usual algebraic properties but failed to reproduce the usual commutation relations. The other one gave the usual result when used on simple fields, but failed to do so for products of fields. Furthermore, it did not fulfill the usual algebraic properties.
We found strong evidence for the statement that twisted NCQFT in a Hamiltonian setting is only consistent if the Hamiltonian is translation invariant.
Finally, we showed that the choice of the interaction term and not the twisting is responsible for the presence or absence of UV/IR-mixing.

These findings are rather discouraging. They indicate that twisted NCQFT is not flexible enough to derive meaningful new predictions. Furthermore, the problem of UV/IR-mixing and the distortion of dispersion relations (and thus violation of Lorentz invariance) is not solved, unless one uses the point-wise product for the interaction.

\subsection*{Acknowledgments}

The author profited a lot from discussions with D.~Bahns, S.~Doplicher, F.~Meyer and in particular K.~Fredenhagen. Financial support from the Graduiertenkolleg ``Zuk\"unftige Entwicklungen in der Teilchenphysik'' is gratefully acknowledged.

\end{document}